\documentclass{svproc}
\usepackage{url}

\usepackage{amsmath,amsfonts,amssymb}
\usepackage{graphicx}
\usepackage{xcolor,color,comment}
\usepackage{natbib}
\numberwithin{equation}{section}
\usepackage[normalem]{ulem}
\usepackage{cancel}

\newcommand{\bs}{\boldsymbol} 
\newcommand{\mb}{\mathbb} 
\newcommand{\mc}{\mathcal} 
\newcommand{\mr}{\mathrm} 
\newcommand{\pd}{\partial}
\newcommand{\wt}{\widetilde} 
\newcommand{\wh}{\widehat} 
\newcommand{\ol}{\overline}
\newcommand{\sub}{\scriptscriptstyle} 
\newcommand{\beq}{\begin{equation}}
\newcommand{\eeq}{\end{equation}}
\newcommand{\beqs}{\begin{subequations}}
\newcommand{\eeqs}{\end{subequations}}
\newcommand{\benum}{\begin{enumerate}[label=(\roman*)]}
\newcommand{\eenum}{\end{enumerate}}

\newcommand{\df}{\mr{d}} 
\newcommand{\Exp}{\mb{E}} 
\newcommand{\alf}{\frac{1}{2}}
\newcommand{\dt}{\df t} 
\newcommand{\dB}{\df B_t} 
 
\newcommand{\dz}{\df z} 
\newcommand{\xvec}{\bs{\mr{x}}}

\newcommand{\uvec}{\bs{\mr{u}}}

\newcommand{\svec}{\bs{\sigma}}
\newcommand{\amat}{\bs{a}}
\newcommand{\tvec}{\bs{\tau}}
\newcommand{\Tvec}{\bs{\mr{T}}}
\newcommand{\sigx}{\svec_{\xvec}}
\newcommand{\sigz}{\sigma_z}

\newcommand{\axz}{\amat_{\xvec z}}
\newcommand{\azz}{a_{zz}}
\newcommand{\atzz}{\wt{a}_{zz}}

\newcommand{\Uvec}{\bs{\mr{U}}}
\newcommand{\intdom}{\mathring{\mr{\Omega}}}

\newcommand{\LL}[1]{{\color{black} #1}}

\begin{document}
\mainmatter              
\title{A Stochastic Ekman--Stokes Model for Coupled 
\LL{Ocean--Wave--Atmosphere} Dynamics}
\titlerunning{Stochastic Coupled Ekman--Stokes Model (SCESM)}  
%
\author{Long Li\inst{1,2} \and Etienne M\'emin\inst{1,2} \and Bertrand Chapron\inst{1,3}} 
\authorrunning{L. Li et al.} 
%
\tocauthor{Long Li, Etienne M\'emin, Bertrand Chapron}
\institute{ODYSSEY Team, Centre Inria de l'Universit\'e de Rennes, France\\
\and
IRMAR – UMR CNRS 6625, Rennes, France\\
\and
Laboratoire d’Oc\'eanographie Physique et Spatiale, Ifremer, Plouzan\'e, France\\
Corresponding author: Long Li, \email{long.li@inria.fr}
}

\maketitle

\begin{abstract}
Accurate representation of atmosphere-ocean boundary layers, including the interplay of turbulence, surface waves, and air--sea fluxes, remains a challenge in geophysical fluid dynamics, particularly for climate simulations. This study introduces a stochastic coupled Ekman--Stokes model (SCESM) developed within the physically consistent Location Uncertainty framework, explicitly incorporating random turbulent fluctuations and surface wave effects. The SCESM integrates established parameterizations for air--sea fluxes, turbulent viscosity, and Stokes drift, and its performance is rigorously assessed through ensemble simulations 
\LL{compared against observations from the LOTUS field experiment}. A performance ranking analysis quantifies the impact of different model components, highlighting the critical role of explicit uncertainty representation in both oceanic and atmospheric dynamics for accurately capturing system variability. 
\LL{Among the tested configurations, the full model version---including both Stokes drift and wave-induced mixing---shows the best agreement with observations.} 
Wave-induced mixing terms improve model performance, while wave-dependent surface roughness enhances air--sea fluxes but reduces the relative influence of wave-driven mixing. This fully coupled stochastic framework provides a foundation for advancing boundary layer parameterizations in large-scale climate models.
\end{abstract}


\section{Introduction}
Numerical modeling of geophysical fluid dynamics, particularly in climate systems, presents significant challenges due to the multi-scale nature of these systems and the complex nonlinear interactions that govern their behavior. Accurately simulating large-scale atmospheric and oceanic flows while maintaining computational efficiency requires the careful modeling or parameterization of subgrid-scale processes. Many of these processes are intermittent, nonlinear, and 
random, underscoring the need for innovative approaches to represent uncertainties. As highlighted by \citet{Hasselmann1976stochastic,Majda1999models,Palmer2019stochastic}, stochastic modeling has emerged as a powerful tool for improving the representation of unresolved dynamics, thereby enhancing the accuracy of weather and climate predictions.

In this study, we adopt a physically consistent stochastic framework known as modeling under location uncertainty (LU), originally introduced by \citet{Memin2014fluid}. This approach preserves fundamental physical properties, such as energy conservation \citep{Bauer2020deciphering}, while naturally incorporating modeling errors and approximations to improve the representation of unresolved scales and their effects. A hierarchy of stochastic geophysical models has been systematically developed and validated within this framework, effectively capturing mesoscale and submesoscale dynamics within large-scale atmospheric and oceanic flows. Examples include the shallow water model \citep{Brecht2021rotating}, the quasi-geostrophic model \citep{Li2023stochastic}, and the primitive hydrostatic model \citep{Tucciarone2025derivation}. By integrating stochasticity, LU-based models improve uncertainty representation and enable coarse-grid simulations to reproduce the long-term statistical properties of high-resolution reference flows.

The LU framework also advances stochastic formulations for wave--current and air--sea interactions by capturing the impact of unresolved small-scale fluctuations on large-scale dynamics. \citet{Bauer2020deciphering} demonstrates that wave-current interactions, including the Coriolis--Stokes and vortex forces in the Craik--Leibovich system, can be interpreted as statistical effects of unresolved flow inhomogeneity within a stochastic flow representation. Building on this, 
\LL{
\citet{Li2025generalized} propose a stochastic formulation of the Ekman--Stokes layer (also referred to as the Wavy Ekman Layer in previous studies \citep[e.g.,][]{McWilliams2012wavy}), which explicitly incorporates spatial variations of the unresolved velocity field. This allows for a more realistic and dynamic response to Stokes vortex forces, which---when spatial variations are neglected---reduce to a quasi-hydrostatic background perturbation. However, much of the dynamical impact of these forces emerges precisely from their spatial structure, as highlighted in \citet{McWilliams2013oceanic, Suzuki2016understanding}. Numerical results in \citet{Li2025generalized} show that the proposed stochastic model produces
}
greater variability, increased kinetic energy, and more extreme events compared to standard models. Sensitivity analyses in that study highlight the influence of transient winds and surface waves, which enhance the dispersion of realizations while maintaining a balanced representation of errors. However, the model treats wind as a prescribed Gaussian process, neglecting its interaction with currents and waves and thereby failing to capture wave-modulated feedback on wind stress.

Expanding upon this foundation, the present work develops a fully coupled stochastic model that integrates turbulence, surface wave effects, and air--sea fluxes, explicitly capturing the interactions between random wind, waves, and currents. This one-dimensional (1D) vertical modeling approach is essential for realistic coarse-scale climate simulations, as it provides a physically consistent framework for parameterizing unresolved boundary layer turbulence and generating vertical profiles at the model's resolution grid points.
\LL{
In this context, we adopt classical bulk flux parameterizations \citep{Fairall1996bulk,Fairall2003bulk} to represent turbulent air–-sea exchanges. These schemes remain widely used in climate models due to their empirical foundation and computational simplicity. While originally developed for deterministic, time-averaged conditions, we apply them here within a stochastic framework as an initial step toward extending traditional parameterizations to account for intrinsic variability. This provides a practical baseline for future developments involving more physically consistent coupling strategies.
To evaluate the performance of the stochastic configurations, we compare model outputs against current profile observations from the LOTUS (Long-Term Upper-Ocean Study) experiment \citep{Price1987wind}. This dataset provides high-quality Eulerian current measurements that are minimally affected by mooring motion, making it a robust benchmark for model validation. LOTUS data have been widely used in the literature for evaluating upper-ocean boundary layer processes \citep{Large1994oceanic,Price1999stratified,Lewis2004time,Polton2005role}.
}

The paper is structured as follows: Section~\ref{sec:models} describes the stochastic coupled Ekman--Stokes model (SCESM) and its parameterizations. Section~\ref{sec:results} presents numerical results and comparisons with observational data. Finally, Section~\ref{sec:conclu} concludes the study and discusses potential avenues for future research.


\section{Models}\label{sec:models}
In this section, we introduce the proposed stochastic framework for the coupled ocean-atmosphere Ekman layers, along with well-established parameterizations for air--sea fluxes, turbulent viscosity, and Stokes drift.

\subsection{Stochastic formulation for air--sea coupled Ekman layers}\label{sec:LU}
\begin{figure}[htbp]
\centering
\includegraphics[width=\textwidth]{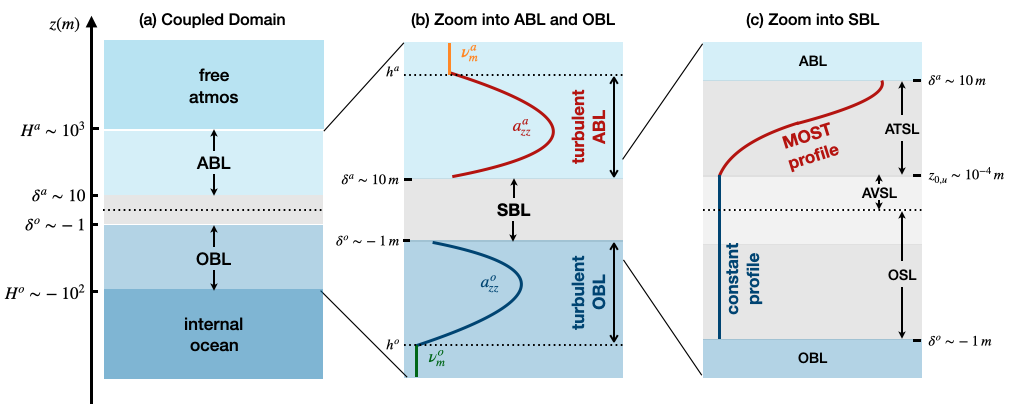}
\caption{\LL{Schematic of the coupled Ekman layers.
(a) Vertical domain structure, including the atmospheric and oceanic boundary layers (ABL and OBL), and their near-surface sublayers.
(b) Turbulent viscosity profiles in both ABL and OBL, defined from the near-surface depths $\delta^a$ and $\delta^o$.
(c) Classical one-sided bulk flux formulation \citep{Pelletier2021twoside} used for surface exchange computations within the surface boundary layer (SBL), which includes the ocean surface layer (OSL), atmospheric viscous sublayer (AVSL), and atmospheric turbulent sublayer (ATSL).}}
\label{fig:illu-model}
\end{figure}

As illustrated in Fig.~\ref{fig:illu-model}, the vertical domain is decomposed as follows: the atmospheric boundary layer (ABL) is defined in $\mr{\Omega}^a = [\delta^a, H^a]$ (with $\delta^a \sim 10$m and $H^a \sim 1000$m), the ocean boundary layer (OBL) is defined in $\mr{\Omega}^o = [H^o, \delta^o]$ (with $\delta^o \sim -1$m and $H^o \sim -100$m), while the air--sea interface consists of the atmospheric surface boundary layer (ASBL) in $[0, \delta^a]$ and the ocean surface boundary layer (OSBL) in $[\delta^o, 0]$.

Following the derivation of the generalized stochastic Craik--Leibovich equations \citep{Bauer2020deciphering} and the generalized stochastic Ekman--Stokes model \citep{Li2025generalized}, we consider coupling two time-dependent Ekman 
\LL{layers for the atmosphere and ocean}. Hereafter, this will be referred to as the stochastic coupled Ekman--Stokes model (SCESM), described by the following stochastic partial differential equations:
\beqs\label{eqs:spde}
\begin{align}\label{eq:spde}
\df \uvec^\alpha 
&= \Big( - i f \big( \uvec^\alpha  + \uvec_s^\alpha - \uvec_g^\alpha \big) + \pd_z \big( \nu^\alpha \pd_z (\uvec^\alpha  + \uvec_s^\alpha) \big) \Big)\, \dt \nonumber \\
&- \Big( \sigz^\alpha \pd_z \big( \uvec^\alpha + \uvec_s^\alpha \big) + i f \sigx^\alpha \Big)\, \dB^\alpha,\ \quad z \in \intdom^\alpha,\ t > 0, 
\end{align}
\beq\label{eq:ic}
\uvec^\alpha (z,0) = \uvec_g^\alpha,\ \quad z \in \intdom^\alpha,
\eeq
\beq\label{eq:bc}
\uvec^\alpha (H^\alpha, t) = \uvec_g^\alpha,\ \quad t > 0,
\eeq
\beq\label{eq:sbc}
\rho^\alpha \nu^\alpha \pd_z \uvec^\alpha (\delta^\alpha, t) = \tvec (t),\ \quad t > 0 .
\eeq
\eeqs
Here, $\alpha \in \{a, o\}$ denotes atmospheric and oceanic components, respectively. The variable $\uvec^\alpha (z,t) = u^\alpha (z,t) + i v^\alpha (z,t)$ represents the horizontal velocity in complex notation, with $i$ as the imaginary unit. 
The Coriolis frequency is denoted by $f$. The viscosity coefficient is given by $\nu^\alpha (z,t) = \nu_m^\alpha + \azz^\alpha (z,t)$, incorporating both molecular and turbulent effects. The uniform density is denoted by $\rho^\alpha$, and $\tvec (t) = \tau_x (t) + i \tau_y (t)$ is the time-dependent surface wind stress.
\LL{
The geostrophic velocity component $\uvec_g^\alpha = u_g^\alpha + i v_g^\alpha$ is assumed time- and depth-independent.
This simplifying assumption is consistent with the reduced 1D formulation of the Ekman boundary layer. However, it prevents the model from capturing potential coupling between geostrophic flow and Stokes drift, a mechanism that has been shown to influence dynamics at submesoscales and, more weakly, at mesoscales \citep{McWilliams2013oceanic, Suzuki2016understanding}. Such interactions, involving the interplay between geostrophic currents, internal gravity waves, and Stokes vortex forces, are important components of the full ocean boundary layer system. Their exclusion represents a limitation of the present formulation, which could be addressed in future extensions.
}

In Eq.~\eqref{eq:spde}, the term $\svec^\alpha\, \dB^\alpha = (\sigx^\alpha\, \dB^\alpha, \sigz^\alpha\, \dB^\alpha)$ represents a stochastic noise component originating from the Location Uncertainty (LU) framework \citep{Memin2014fluid,Resseguier-GAFD-17,Bauer2020deciphering}, which accounts for unresolved turbulent motions probabilistically. 
The noise is spatially correlated through the correlation operator $\svec^\alpha$, acting on a cylindrical Brownian motion $B_t^\alpha$. 
\LL{In this framework, the It\^o quadratic variation process \citep{Bauer2020deciphering} associated with the vertical noise component is defined as
}
\beqs\label{eqs:noise-stat}
\beq
\alf \df \left\langle \int_0^\bullet \sigz^\alpha\, \df B_s^\alpha, \int_0^\bullet \sigz^\alpha\, \df B_s^\alpha\right\rangle_t = \alf \sigz^\alpha \sigz^\alpha\, \dt =: \azz^\alpha\, \dt .
\eeq
\LL{Physically, \(\azz^\alpha\) plays the role of an eddy viscosity induced by unresolved turbulent fluctuations.}
The It\^o--Stokes drift (introduced by \citet{Bauer2020deciphering}), $\uvec_s = \pd_z \axz$ (within the Ekman layer scalings \citep{Li2025generalized}), captures the vertical inhomogeneity of co-quadratic variation between horizontal and vertical noise components:
\beq
\alf \df \left\langle \int_0^\bullet \sigx^\alpha\, \df B_s^\alpha, \int_0^\bullet \sigz^\alpha\, \df B_s^\alpha\right\rangle_t = \alf \sigx^\alpha \sigz^\alpha\, \dt =: \axz^\alpha\, \dt .
\eeq
\eeqs
\LL{
In this formulation, the unresolved variability in the atmosphere and ocean is modeled through independent (cylindrical) Brownian motions $B_t^a$ and $B_t^o$, representing intrinsic turbulence internal to each medium. This assumption enables a tractable stochastic representation of subgrid-scale processes. However, we acknowledge that in reality, coherent structures such as wave-phase-aligned turbulence or wave-induced surface-layer motions may introduce physical correlations across the interface \citep[e.g.,][]{Qiao2016wave}. Capturing such effects would require a more general framework with correlated noises, which represents an interesting direction for future developments.
}

The above provides a general description of our stochastic model. To specify further the noise, we may also formulate the following inverse problem: determining a specific noise such that the resulting statistical properties correspond to a prescribed turbulent viscosity \LL{$\atzz^\alpha$} and a given Stokes drift \LL{$\wt{\uvec}_s^\alpha$}. 
\LL{
In particular, a simple projection-based construction was proposed by \citet{Li2025generalized} in the ocean-only setting; this approach is adopted here in the coupled atmosphere–-ocean context. 
}

Given the turbulent viscosity \LL{$\atzz^\alpha$}, we adopt the following spectral decomposition for the vertical noise 
component:
\beqs\label{eq:noises}
\beq\label{eq:noise-z}
\sigz^\alpha\, \dB^\alpha 
= \sqrt{2} \sum_{n>0} \big[ (\LL{\atzz^\alpha})^{\sub 1/2}, e_n^\alpha \big] e_n^\alpha\, \df \beta_{t,n}^\alpha ,
\eeq
where $\{e_n^\alpha \LL{(z)}\}_{n>0}$ denotes a set of localized basis functions with minimal overlapping support in the real-valued Hilbert space $L^2 (\mr{\Omega}^\alpha, \mb{R})$, 
equipped with the inner product $[ f, g ] = \int_{\mr{\Omega}^\alpha} f (z) g (z)\, \dz$. The set $\{\beta_{t,n}^\alpha\}_{n>0}$ consists of independent one-dimensional real-valued Brownian motions. Furthermore, the two sets of Brownian motions, $\{\beta_{t,n}^o\}_{n>0}$ and $\{\beta_{t,n}^a\}_{n>0}$, are assumed to be independent. 

Given the anti-derivative of the horizontal Stokes drift, defined as $\LL{\wt{\Uvec}_s^\alpha} (z) = \int_{\sub H^o}^{z} \LL{\wt{\uvec}_s^\alpha} (\zeta)\, \df \zeta$, where \LL{$\wt{\uvec}_s^\alpha (z) = \wt{u}_s^\alpha (z) + i \wt{v}_s^\alpha (z)$}, the horizontal noise component 
can be constructed as
\beq\label{eq:noise-x}
\sigx^\alpha\, \dB^\alpha = \sqrt{2} \sum_{n>0} \big[ (\LL{\atzz^\alpha})^{\sub -1/2} \LL{\wt{\Uvec}_s^\alpha}, e_n^\alpha \big] e_n^\alpha\, \df \beta_{t,n}^\alpha , 
\eeq
\eeqs
where the inner product is \LL{defined} as $[\Uvec, e_n] = [U, e_n] + i [V, e_n]$. 
\LL{
The corresponding  quadratic variation process, defined in Eq.~\eqref{eqs:noise-stat}, are then given by
\beqs
\beq
\azz^\alpha = \sum_n \big[ (\atzz^\alpha)^{\sub 1/2}, e_n^\alpha \big]^2\, (e_n^\alpha)^2 ,
\eeq
\beq
\axz^\alpha = \sum_n \big[ (\atzz^\alpha)^{\sub 1/2}, e_n^\alpha \big] \big[ (\atzz^\alpha)^{\sub -1/2} \wt{\Uvec}_s^\alpha, e_n^\alpha \big]\, (e_n^\alpha)^2 .
\eeq
\eeqs
By Parseval’s theorem, these reconstructed functions are globally consistent with the given inputs $\atzz^\alpha$ and $\wt{\Uvec}_s^\alpha$, in the sense that
\beq
\int_{\mr{\Omega}^\alpha} \azz^\alpha (z)\, \dz 
= \int_{\mr{\Omega}^\alpha} \atzz^\alpha (z)\, \dz ,\ \quad
\int_{\mr{\Omega}^\alpha} \axz^\alpha (z)\, \dz 
= \int_{\mr{\Omega}^\alpha} \wt{\Uvec}_s^\alpha (z)\, \dz ,\ \quad
\eeq
Since the basis functions $e_n$ are assumed to be localized (i.e., significant over small, distinct regions with negligible overlap), Parseval’s identity applies approximately pointwise. This yields $\azz^\alpha (z) \approx \atzz^\alpha (z)$ and $\axz^\alpha (z) \approx \wt{\Uvec}_s^\alpha (z)$, so that the resulting It\^o--Stokes drift approximates the given Stokes drift: $\uvec_s^\alpha = \pd_z \axz^\alpha \approx \wt{\uvec}_s^\alpha$.  
This identification in the oceanic case has been illustrated in \citet{Li2025generalized}. For simplicity, we use the same notation for both the prescribed fields and their associated noise quadratic variations.
}
 
\LL{The Stokes drift is typically} not included in the atmospheric momentum equation \citep{Lewis2004time}, as it is generally much smaller than the geostrophic wind. \LL{Therefore, in} the following, we neglect $\uvec_s^a$ by assuming negligible atmospheric horizontal noise ($\sigx^a\, \dB^a \approx 0$) and refer to the Stokes drift $\uvec_s$ exclusively in the ocean. As a result, the stochastic atmospheric Ekman model \LL{can read}
\beq\label{eq:ua}
\df \uvec^a = \Big( -i f \big( \uvec^a - \uvec_g^a \big) + \pd_z \big( \nu^a \pd_z \uvec^a \big) \Big)\, \dt -  \sigz^a \pd_z \uvec^a\, \dB^a .
\eeq
Taking the expectation, the mean wind velocity $\Exp[\uvec^a]$ satisfies 
\beq
\pd_t \Exp[\uvec^a] = -i f \Big( \Exp[\uvec^a] - \uvec_g^a \Big) + \pd_z  \Big( \Exp \big[ \nu^a \pd_z \uvec^a \big] \Big) ,
\eeq
where the last term is generally nonlinear, as the turbulent viscosity closure for $\azz^a$ (recalling that $\nu^a = \nu_m^a + \azz^a$) typically depends on the pathwise solution $\uvec^a$, at least through air--sea coupling, as will be demonstrated later. The stochastic model \eqref{eq:ua} can therefore be interpreted as a generalization of the classical model, incorporating random fluctuation effects arising from unresolved smaller scales.

Similarly, taking the expectation of Eq.~\eqref{eq:spde} for the ocean component, the mean current velocity $\Exp[\uvec^o]$ satisfies
\beq
\pd_t \Exp[\uvec^o] = -i f \Big( \Exp[\uvec^o] + \Exp[\uvec_s] - \uvec_g^o \Big) + \pd_z  \Big( \Exp \big[ \nu^o \pd_z \uvec^o \big] + \Exp \big[ \nu^o \pd_z \uvec_s \big] \Big) ,
\eeq
which remains a nonlinear equation---because of the dependency of $\mu^{o}$ on $\uvec^o$.  In general, the Stokes drift may also exhibit stochastic variability, as will be shown later. Unlike the classical Ekman--Stokes model, which accounts only for the Coriolis--Stokes force $i f \Exp[\uvec_s]$, the proposed mean model introduces an additional term, $\pd_z \big( \Exp[\nu^o \pd_z \uvec_s] \big)$, representing a \emph{wave mixing} effect. This term initially arises from a change of variables in the derivation of the generalized stochastic Craik--Leibovich equation \citep{Bauer2020deciphering}.

To further examine its influence, one can vertically integrate the mean current velocity. The mean current transport, given by $\Exp[\Tvec^o] = \int_{\sub\intdom^o} \Exp[\uvec^o] (z)\, \dz$, then satisfies
\beq
\pd_t \Exp[\Tvec^o] = -i f \Big( \Exp[\Tvec^o] + \Exp[\Tvec_s] - \Tvec_g^o \Big) + \frac{1}{\rho^o} \Big( \Exp[\tvec] + \Exp[\tvec_s] \Big) ,
\eeq
where $\tvec_s = \rho^o \nu^o \pd_z \uvec_s (\delta^o)$  represents an additional surface wave stress absent in the classical Ekman--Stokes model.

Returning to the general case, the SCESM \eqref{eqs:spde}, combined with the simplified atmospheric equation \eqref{eq:ua} and the specific noise representations \eqref{eq:noises}, establishes a coupled framework for the atmospheric and oceanic boundary layers, incorporating both surface wave effects and random turbulent fluctuations. However, the SCESM does not account for stratification (buoyancy) effects, which will be investigated in future work. To complete the model, we present parameterizations for the air--sea momentum flux $\tvec$, the turbulent viscosity $\azz^\alpha$, and the Stokes drift $\uvec_s$ in the subsequent sections, all of which are based on well-established approaches.

\subsection{Bulk parameterization of turbulent air--sea fluxes}\label{sec:bulk}
In this section, we briefly review the parameterization of turbulent fluxes for the ocean--atmosphere coupling. We begin by recalling the atmospheric surface boundary layer model, which follows the Monin--Obukhov similarity theory (MOST) \citep{Monin1954basic}, a generalized law of the wall for stratified fluids (see 
\LL{\citet{Foken2006MOST}} for a detailed introduction on the subject). 
\LL{In the classical one-sided bulk flux formulation \citep{Pelletier2018sensitivity}, which remains standard in many climate models,}
the resolved wind $\uvec$, potential temperature $\theta$ and humidity $q$ at the atmospheric surface layer depth $\delta^a$, can be expressed in terms of their 
\LL{Monin--Obukhov (M--O) scaling parameters \citep{FoxKemper2022ocean}} $(u_*, \theta_*, q_*)$, as follows:  
\beqs\label{eqs:most}
\beq
\big|[\uvec]_{\sub\delta^o}^{\sub\delta^a}\big| = \frac{u_*}{\kappa} \left( \ln \Big( \frac{\delta^a}{z_{0,u} (u_*, \theta_*, q_*)} \Big) - \psi_m \Big( \frac{\delta^a}{L_O (u_*, \theta_*, q_*)} \Big) \right) ,
\eeq
\beq
[\theta]_{\sub\delta^o}^{\sub\delta^a} = \frac{\theta_*}{\kappa} \left( \ln \Big( \frac{\delta^a}{z_{0,\theta} (u_*, \theta_*, q_*)} \Big) - \psi_h \Big( \frac{\delta^a}{L_O (u_*, \theta_*, q_*)} \Big) \right) ,
\eeq
\beq
[q]_{\sub\delta^o}^{\sub\delta^a} = \frac{q_*}{\kappa} \left( \ln \Big( \frac{\delta^a}{z_{0,q} (u_*, \theta_*, q_*)} \Big) - \psi_h \Big( \frac{\delta^a}{L_O (u_*, \theta_*, q_*)} \Big) \right) ,
\eeq
\eeqs
where $[X]_{\sub\delta^o}^{\sub\delta^a} = X (\delta^a) - X (\delta^o)$
denotes the sea-surface-relative (under the ocean constant profiles assumption) value of the atmospheric quantity $X$, while $\kappa \approx 0.4$ is the von K\'arm\'an constant, $z_{0,u}$, $z_{0,\theta}$ and $z_{0,q}$ represent the surface roughness lengths for momentum, heat, and humidity, respectively. These roughness lengths depend on the 
\LL{scaling parameters $(u_*, \theta_*, q_*)$}. Additionally, $L_O$ is the Obukhov length, which also depends on the 
\LL{scaling parameters}, and $\psi_m$ and $\psi_h$ are the stability functions for momentum and tracer. \LL{Explicit} forms of these functions can be found in \citet{Beljaars1991flux,Grachev2000convective}. For the neutral case (without stratification), the above relationships reduce to the classical law of the wall, which is simply a logarithmic profile. 

\LL{
Note that the classical one-sided bulk formulation \eqref{eqs:most} assumes surface values based on near-surface measurements or model outputs, effectively treating the ocean surface layer as vertically uniform. While this simplification was originally introduced for computational efficiency in coarse-resolution models, it can introduce biases in the near-surface structure---particularly when surface-intensified processes such as Stokes shear are active within the upper meter. A two-sided surface-layer parameterization was proposed by \citet{Pelletier2021twoside} to address this issue. With modern ocean models now resolving the upper ocean boundary layer at vertical scales of order 1 m or finer, and with increasing recognition of the role of near-surface dynamics, the adoption of more consistent formulations \citep[e.g.,][]{Large2019similarity,FoxKemper2022ocean} provides a promising direction for future improvements.
}

The Obukhov length $L_O$ depends on the friction velocity scale $u_*$ and the atmospheric buoyancy flux scale $B_f$ as follows:
\beq\label{eq:Lo}
L_O = \frac{u_*^3}{\kappa B_f},\ \quad B_f = g u_* \left( \frac{\theta_*}{\theta_v (\delta_a)} + \frac{q_*}{q (\delta_a) + q_0} \right) ,
\eeq
where $g \approx 9.81$ m s$^{-2}$ is the gravity acceleration, $\theta_v (z) = \theta (z) \big( 1 + q(z) / q_0 \big)$ is the virtual potential temperature, and $q_0 \approx 0.61$ kg kg$^{-1}$ is the specific humidity of saturated air. 

Equations \eqref{eqs:most} are nonlinear, depending on the unknown 
\LL{scaling parameters} $(u_*, \theta_*, q_*)$. To solve these three variables, a fixed-point iterative algorithm can be used. This numerical counterpart of the nonlinear formulation is referred to as the ``bulk formula''. Various versions exist, differing mainly in the initialization of the algorithm, the parameterization of surface roughness lengths \LL{and the choice of stability functions}. In this study, we adopt the well-established COARE (Coupled Ocean--Atmosphere Response Experiment) algorithm \citep{Fairall1996bulk,Fairall2003bulk}.

\LL{The} wind stress $\tvec$, the sensible heat flux $Q_{SH}$, and the latent heat flux $Q_{LH}$ across the air-sea interface are given by     
\beqs\label{eqs:flux}
\beq
|\tvec| = \rho^a C_d U \big|[\uvec]_{\sub\delta^o}^{\sub\delta^a}\big| = \rho^a u_*^2 \big|[\uvec]_{\sub\delta^o}^{\sub\delta^a}\big| / U , 
\eeq
\beq
Q_{SH} = \rho^a c_P C_h U [\theta]_{\sub\delta^o}^{\sub\delta^a} = \rho^a c_P u_* \theta_* ,
\eeq
\beq
Q_{LH} = \rho^a l_E C_e U [q]_{\sub\delta^o}^{\sub\delta^a} = \rho^a l_E u_* q_* ,
\eeq
\eeqs
where $c_P$ is the specific heat of air and $l_E$ is the latent heat of vaporization. The transfer coefficients $(C_d, C_h, C_e)$ for momentum, heat and humidity all depend on 
$(u_*, \theta_*, q_*)$, and are determined from \LL{\eqref{eq:Lo} and \eqref{eqs:flux}}:
\beqs
\beq
C_d = \left( \frac{\kappa}{\ln (\delta_a / z_{0,u}) - \psi_m (\delta_a / L)} \right)^2 ,
\eeq
\beq
C_h = \frac{\kappa C_d^{1/2} 
(\delta_a)}{\ln (\delta_a / z_{0,\theta}) - \psi_h (\delta_a / L)} ,
\eeq
\beq
C_e = \frac{\kappa C_d^{1/2} 
(\delta_a)}{\ln (\delta_a / z_{0,q}) - \psi_h (\delta_a / L)} .
\eeq
\eeqs
In Eqs.~\eqref{eqs:flux}, $U$ denotes the scale the scalar difference in velocity across the air--sea interface, incorporating a ``gustiness'' factor $u_{\text{gust}}$, defined as
\beqs
\beq
U = \left( \big|[\uvec]_{\sub\delta^o}^{\sub\delta^a}\big|^2 + u_{\text{gust}}^2 \right)^{1/2} ,
\eeq
\beq
u_{\text{gust}} = 
\left\{
\begin{aligned}
&1.2 (B_f z_i)^{1/3}, && \text{if} \ B_f > 0 \\
&0.2, && \text{if} \ B_f \leq 0 ,
\end{aligned}
\right.
\eeq
\eeqs
where $z_i \approx 600$ m is the convective boundary layer depth. The gustiness parameter accounts for atmospheric stability and is included in the COARE algorithm to ensure non-zero momentum fluxes at low wind speeds, by parameterizing the convective effect on momentum transfer. 

Finally, the surface roughness lengths are parameterized in COARE by
\beqs
\beq\label{eq:rough-m}
z_{0,u} = \alpha_{ch} \frac{u_*^2}{g} + 0.11 \frac{\nu_m^a}{u_*} ,
\eeq
\beq
z_{0,\theta} = \min \left( 1.15\times10^{-4}, 5.5\times10^{-5} R_r^{-0.6} \right),\ \quad R_r = z_{0,u} \frac{u_*}{\nu_m^a} ,
\eeq
\beq
z_{0,\theta} = z_{0,q} .
\eeq
\eeqs
The momentum roughness in \eqref{eq:rough-m} is separated into a rough-flow component using Charnock scaling \citep{Charnock1955wind} and a smooth-flow component with a fixed roughness Reynolds number. The Charnock coefficient $\alpha_{ch}$ can be parameterized in different ways to account for physical effects. One such approach, based on wind speed \citep{Fairall2003bulk}, uses a piecewise increasing and affine function of $U$:
\beqs
\beq\label{eq:wind-speed}
\alpha_{ch} (U) = 
\left\{
\begin{aligned}
&0.011, && \text{if} \ U \leq 10\, \text{m s}^{-1} \\
&0.011 + 0.007 (U - 10) / 8,  && \text{if} \ U \in (10,18)\, \text{m s}^{-1} \\
&0.018, && \text{if} \ U \geq 18\, \text{m s}^{-1} .
\end{aligned}
\right.
\eeq
Alternatively, the Charnock coefficient can be parameterized as a function of wave age \citep{Maat1991roughness}:
\beq\label{eq:wave-age}
\alpha_{ch} (u_*, C_p) = A \left(\frac{u_*}{C_p}\right)^B ,
\eeq
where $C_p$ is the phase speed of the waves at the spectral peak. The constants $A = 0.114$ and $B = 0.622$ are used in COARE 3.5 \citep{Edson2013exchange}.

Another approach is to parameterize surface roughness based on wave slope, as in the sea state– and wave age–dependent formulation \citep{Donelan1990airsea}:
\beq\label{eq:sea-state}
\alpha_{ch} (u_*, C_p, H_s) = A' H_s \left(\frac{u_*}{C_p}\right)^{B'} \frac{g}{u_*^2} ,
\eeq
where $H_s$ is the significant wave height. The constants $A’ = 0.091$ and $B’ = 2$ are used in COARE 3.5 \citep{Edson2013exchange}.
\eeqs

In this work, we employ the COARE (fixed-point) algorithm, using the relative wind
$[\uvec]_{\sub\delta^o}^{\sub\delta^a} (t)$
at each time step, 
\LL{while assuming constant stratification: the potential temperature difference $[\theta]_{\sub\delta^o}^{\sub\delta^a}$ and specific humidity difference $[q]_{\sub\delta^o}^{\sub\delta^a}$ are kept fixed throughout the simulation.}
This allows us to first determine the  
\LL{scaling parameters} $(u_*, \theta_*, q_*)$ along with the intermediate transfer coefficients $(C_d, C_h, C_e)$. Consequently, the wind stress is given by $\tvec (t) = \rho^a \big( u_*^2 [\uvec]_{\sub\delta^o}^{\sub\delta^a} / U \big) (t)$, which provides the surface boundary condition \eqref{eq:sbc} at each time step. This formulation assumes a stratified atmospheric surface layer, albeit with steady or time-averaged tracers, while considering the neutral SCESM \eqref{eqs:spde}. The heat fluxes in \eqref{eqs:flux} will be incorporated in future investigations of stratified Ekman boundary layers, particularly when incorporating the evolution equations for temperature \citep{McWilliams2009buoyancy}.

\LL{
We remark that the COARE bulk algorithm is derived from time- and space-averaged observations and effectively parameterizes complex near-surface processes, including wave, spray, and bubble dynamics. In our stochastic framework, it is applied pathwise to the resolved variables, allowing stochastic fluctuations to influence the fluxes through time integration. This approach implies an instantaneous adjustment of the air–sea exchange layers to stochastic inputs---an assumption that may overlook the finite response times and nonlocal effects associated with wave-mediated momentum transfer. While practical and consistent with current coarse-resolution coupled models, this assumption could be relaxed in future work by incorporating stochastic extensions of similarity theory or asynchronous coupling strategies \citep{Valcke2021coupling}.
}

\subsection{Turbulent viscosity closure and Stokes drift}\label{sec:kpp-stokes}
In this work, we implement the diagnostic K-Profile Parameterization (KPP) scheme \citep{Large1994oceanic} for turbulent viscosity closure. \LL{While the classical bulk flux formulation (Fig.~\ref{fig:illu-model}c) applies MOST extrapolation only on the atmospheric side, the turbulent viscosity in each domain is computed from KPP with M--O surface scaling. This ensures consistency in turbulent closure across both fluids, as shown schematically in Fig.~\ref{fig:illu-model}b.}

The KPP scheme is formulated in a nondimensional vertical coordinate $\zeta^\alpha = |z| / h^\alpha$, where $h^\alpha$ represents the turbulent boundary layer depth. The general formulation is expressed as:
\beq
\azz^\alpha (\zeta^\alpha) = h^\alpha w (\zeta^\alpha) G (\zeta^\alpha),\ \quad w (\zeta^\alpha) = \frac{\kappa u_*}{\phi_m (\zeta h^\alpha / L_O)} ,
\eeq
where $w$ is the turbulent vertical velocity scale and $G$ is a fourth-order polynomial \citep{Large1994oceanic}, ensuring that the viscosity and its gradient match specific values at the top and bottom of the boundary layer. 
The universal function $\phi_m$ is set to unity for neutral conditions, simplifying the formulation to match the profiles of \citet{OBrien1970note}.

In the stratified case, the boundary layer depth $h^\alpha$ depends on both velocity and buoyancy, with the bulk Richardson number \citep{Large1994oceanic} governing the relationship. In the particular case of neutral conditions, the Ekman layer depth is defined by:
\beq
h^\alpha (t) = \frac{c^\alpha}{|f|} u_* (t) ,
\eeq
where the atmospheric constant $c^a = 0.2$ \citet{Arya1981param} and the oceanic constant $c^o = -0.7$ \citet{Large1994oceanic} are used. These constants are consistent with MOST.

It is worth noting that a prognostic turbulence scheme, based on the generic length scale theory \citep{Umlauf2003generic}, and specifically second-moment closure \citep{Harcourt2013second, Harcourt2015improved} for Langmuir turbulence, which solves a turbulent kinetic energy (TKE) equation, could be explored in future studies following a similar framework.

In general, surface gravity waves exhibit a broad spectrum, leading to a complex vertical profile for the Stokes drift \citep{Huang1971derivation, Jenkins1989use}. For simplicity, we consider a steady, monochromatic deep-water wave in this study. The surface elevation, accurate to the leading order in wave steepness, is expressed as $\eta = \eta_0 \cos (k x - \omega t)$, where $\eta_0$ is the wave amplitude, $k$ is the horizontal wavenumber, and $\omega = (g k)^{1/2}$ is the angular frequency, consistent with the deep-water dispersion relation. The corresponding horizontal components of the Stokes drift are approximated by \citet{Phillips1977dynamics} as:
\begin{equation}\label{eq:stokes}
\uvec_s (z) = U_s e^{2 k z} e^{i \theta_s} ,  
\end{equation} 
where $U_s = \omega k \eta_0^2$ is the magnitude of the Stokes drift, and $\theta_s$ represents the wave propagation direction. 
Following \citet{Li2025generalized}, the wave direction $\theta_s$ is parameterized by a normal distribution: $\theta_s \sim \mc{N} (\Theta_s, \Sigma_s)$, where the mean direction $\Theta_s$ is aligned with the geostrophic wind, and the small standard deviation $\Sigma_s$ represents the angular uncertainty, accounting for the misalignment between wind and wave direction.

Note that the anti-derivative $\Uvec_s$, used in Eq.\eqref{eq:noise-x} for the Stokes drift profile, is given by $\uvec_s / (2 k)$. For the monochromatic wave described in Eq.\eqref{eq:stokes}, the phase speed of the waves at the spectral peak and the significant wave height, used in the wave-age dependent formulation \eqref{eq:wave-age} and \eqref{eq:sea-state} for surface roughness parameterization, are defined as \citep{McWilliams2014Langmuir}:
\beq
C_p = \omega/k = \sqrt{g/k},\ \quad H_s = 2 \sqrt{2} \eta_0 .
\eeq


\section{Results}\label{sec:results}
In this section, we numerically investigate the proposed stochastic coupled Ekman-Stokes model (SCESM) with the presented parameterization through ensemble simulations.

\subsection{Model configurations}
The configuration parameters are presented in Table~\ref{tab:parameters}, with most values selected to be consistent with the observational data discussed later.
The SCESM is solved numerically using a pseudo-spectral Chebyshev method combined with an implicit time-stepping scheme \citep{Li2025generalized}. The ocean and atmosphere domains are discretized with 300 and 1000 Chebyshev nodes, respectively, and 500 ensemble members are used. 
\LL{The coupled model is integrated over 20 days using a uniform time step of $\Delta t^a = \Delta t^o = 300$ s. In this configuration, air–sea coupling is performed at each time step using the instantaneous surface wind and current. If a larger ocean time step were employed relative to the atmosphere, asynchronous or synchronous coupling strategies could be considered \citep{Valcke2021coupling}, and further improvements may be achieved by incorporating the Schwarz iterative method \citep{Marti2021Schwarz}.}
\begin{table*}[htbp]
    \centering
    \caption{Model Parameters}
    \begin{tabular}{c c l}
        \hline
        \textbf{Symbol} & \textbf{Value} & \textbf{Description} \\
        \hline
        $H^a$ & $1000$ m & Upper bound of atmospheric domain \\
        $H^o$ & $-100$ m & Lower bound of oceanic domain \\
        $\delta^a$ & $10$ m & Lower bound of atmospheric domain \\
        $\delta^o$ & $-1$ m & Upper bound of oceanic domain \\
        $f$ & $8.36 \times 10^{-5}$ s$^{-1}$ & Coriolis parameter \\
        $\nu_m^a$ & $1.5 \times 10^{-5}$ m$^2$ s$^{-1}$ & Air kinematic viscosity \\
        $\nu_m^o$ & $10^{-6}$ m$^2$ s$^{-1}$ & Water kinematic viscosity \\
        $\rho^o$ & $10^3$ kg m$^{-3}$ & Water density \\
        $\rho^a$ & $1$ kg m$^{-3}$ & Air density \\
        $\uvec_g^a$ & $(9 + 0i)$ m s$^{-1}$ & Geostrophic wind velocity \\
        $\uvec_g^o$ & $0$ m s$^{-1}$ & Geostrophic current velocity \\
        $\theta^a (\delta^a)$ & $299.65$ K (26.5$^\circ$C) & Atmospheric potential temperature at $\delta^a$ \\
        $\theta^o (\delta^o)$ & $301.15$ K (28$^\circ$C) & Oceanic potential temperature at $\delta^o$ \\
        $q (\delta^a)$ & $0\%$ & Atmospheric specific humidity at $\delta^a$ \\
        $\eta_0$ & $0.8$ m & Surface wave amplitude \\
        $\lambda$ & $60$ m & Wavelength of surface wave \\
        $\Theta_s$ & $0^\circ$ & Wave mean propagation direction \\
        $\Sigma_s$ & $5^\circ$ & Wave spreading angle \\
        \hline
    \end{tabular}
    \label{tab:parameters}
\end{table*}

To assess the contributions of different modeling terms in the SCESM, we compare it against several reduced versions:
\begin{itemize}
    \item \textbf{RAM}: Stochastic atmosphere model coupled with a deterministic ocean model without Stokes drift:
    \beqs
    \begin{align}
    \df \uvec^a &= \Big( -i f \big( \uvec^a - \uvec_g^a \big) + \pd_z \big( \nu^a \pd_z \uvec^a \big) \Big)\, \dt - \sigz^a \pd_z \uvec^a\, \dB^a , \label{eq:ua-spde}\\
    \pd_t \uvec^o &= -i f \big( \uvec^o - \uvec_g^o \big) + \pd_z \big( \nu^o \pd_z \uvec^o \big) .
    \label{eq:uo-pde}
    \end{align}
    \eeqs
    \item \textbf{ROM}: Deterministic atmosphere model coupled with a stochastic ocean model without Stokes drift:
    \beqs
    \begin{align}
    \pd_t \uvec^a &= -i f \big( \uvec^a - \uvec_g^a \big) + \pd_z \big( \nu^a \pd_z \uvec^a \big) , \label{eq:ua-pde}\\
    \df \uvec^o &= \Big( -i f \big( \uvec^o - \uvec_g^o \big) + \pd_z \big( \nu^o \pd_z \uvec^o \big) \Big)\, \dt - \sigz^o \pd_z \uvec^o\, \dB^o . 
    \label{eq:uo-spde}
    \end{align}
    \eeqs
    \item \textbf{RCM}: Stochastic atmosphere model \eqref{eq:ua-spde} coupled with a stochastic ocean model \eqref{eq:uo-spde} without Stokes drift.
    \item \textbf{RCM-RS}: RCM with Stokes drift under random angular directions, but without wave mixing terms. That is, Eq.~\eqref{eq:ua-spde} with
    \beq\label{eq:RCM-RS}
    \df \uvec^o = \Big( -i f \big( \uvec^o + \uvec_s - \uvec_g^o \big) + \pd_z \big( \nu^o \pd_z \uvec^o \big) \Big)\, \dt - \Big( \sigz^o \pd_z \uvec^o + i f \sigx^o \Big)\, \dB^o .
    \eeq
    \item \textbf{RCM-RS-WM}:  RCM-RS with wave mixing terms. That is, Eq.~\eqref{eq:ua-spde} with
    \begin{align}\label{eq:RCM-RS-WM}
    \df \uvec^o &= \Big( -i f \big( \uvec^o + \uvec_s - \uvec_g^o \big) + \pd_z \big( \nu^o \pd_z (\uvec^o + \uvec_s) \big) \Big)\, \dt \nonumber \\
    &- \Big( \sigz^o \pd_z \big( \uvec^o + \uvec_s \big) + i f \sigx^o \Big)\, \dB^o .
    \end{align}
\end{itemize}
All models are initialized with the same initial condition \eqref{eq:ic} and use the same Dirichlet boundary condition \eqref{eq:bc}.
\LL{
We remark that Equations \eqref{eq:RCM-RS} and \eqref{eq:RCM-RS-WM} span two simplified configurations---excluding or including Stokes shear in the turbulent boundary condition---to explore the sensitivity of the coupled system to different assumptions. In reality, the interplay between turbulence and wave-induced effects is more intricate. Turbulence tends to reduce the Lagrangian shear \citep{Pearson2018turbulence}, and adjustments such as the Coriolis--Stokes interaction give rise to the so-called ``Lagrangian Thermal Wind'' balance \citep{McWilliams2013oceanic,Haney2015symmetric,Suzuki2016understanding}. These processes can produce an ``anti-Stokes'' Eulerian shear that opposes the Stokes drift, resulting in a small net Lagrangian shear. While the COARE algorithm assumes that Eulerian velocity enters the surface flux formulation, the Eulerian current is not independent of wave effects---such as vortex, Coriolis-–Stokes, and advection feedbacks---raising fundamental questions about whether Eulerian or Lagrangian formulations provide a better basis for bulk parameterization. This is an open debate in the literature \citep[see][]{Vanneste2022Stokes}, and our two configurations are intended as limiting-case models within this ongoing discussion. A more comprehensive treatment involving wave-averaged closures and consistent shear-dependent fluxes is left for future work.
}

To distinguish the impact of different modeling terms from that of parameterization choices, we first conduct a comparative analysis using the same wind speed-dependent formulation \eqref{eq:wind-speed} for the surface roughness length. This ensures that the air--sea flux condition \eqref{eq:sbc} remains consistent across models.

\begin{figure}[htbp]
\centering
\includegraphics[width=\textwidth]{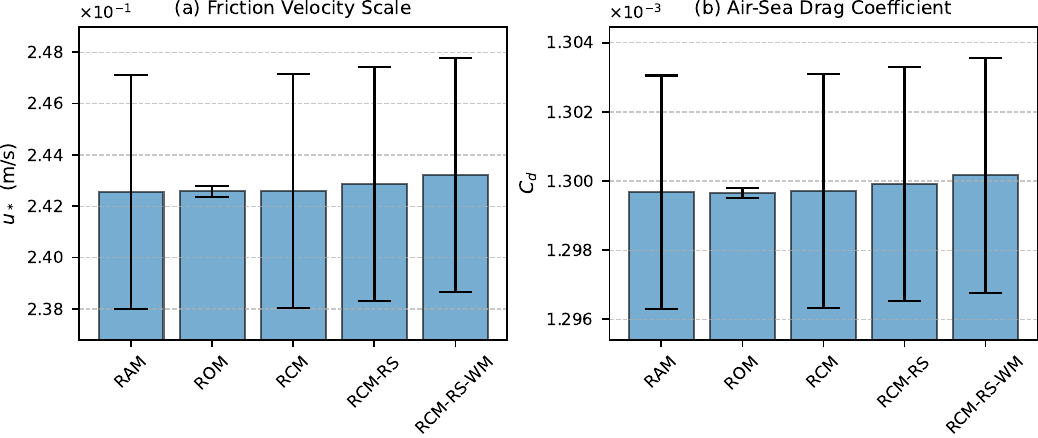}
\par\medskip
\caption{Comparison of the ensemble mean (represented by bars) and spread (represented by error bars) for (a) friction velocity $u_*$ and (b) air-sea momentum transfer coefficient $C_d$, 
across different random models (labeled on the x-axis). The ensemble mean and spread ($\mr{mean} (u) \pm \mr{std} (u)$) are time-averaged over the last 10 days.}
\label{fig:diag}
\end{figure}

For example, Fig.~\ref{fig:diag} demonstrates that all models yield similar ensemble mean values for the friction velocity $u_*$ and the air--sea momentum transfer coefficient $C_d$, with RCM-RS and RCM-RS-WM exhibiting slight increases in these values. Additionally, the overall level of uncertainty is similar across models, except for ROM, which explicitly incorporates stochasticity only in the ocean. Unsurprisingly, this significantly reduces the uncertainty representation of air--sea fluxes.

In this context, we first compare the model performances against observations in Section~\ref{sec:lotus}, followed by a statistical analysis of model diagnostics in Section~\ref{sec:stats}. Additional results incorporating wave-dependent formulations \eqref{eq:wave-age} and \eqref{eq:sea-state} for RCM-RS and RCM-RS-WM are presented in Section~\ref{sec:wave}.

\subsection{Comparison with LOTUS observations}\label{sec:lotus}
To assess the models' performance, we compare their outputs with observations from the Long-Term Upper-Ocean Study (LOTUS) experiment \citep{Price1987wind,Price1999stratified}, conducted in the western Sargasso Sea ($34^\circ$N, $70^\circ$W) during the summer of 1982. This dataset provides 160 days of current profile measurements collected using vector-measuring instruments mounted on a stable platform to mitigate errors associated with mooring motion. The geostrophic velocity, assumed constant at a depth of 50 m, was removed to extract the wind-driven signal. Wind and current data were daily averaged to reduce high-frequency oscillations in the observed signals, rotated to align the wind direction with nominal north, and subsequently averaged over the 160-day period. Table 1 in \citet{Price1987wind} reports the mean values and confidence intervals for the downwind and crosswind current components at depths $z=(-5, -10, -15, -25)$ m.

To ensure comparability with observations, the model's eastward and northward \LL{(Eulerian)} ocean velocity components ($u^o$ and $v^o$) on the wind-relative coordinates  are rotated into a wind-relative coordinate system. The downwind and crosswind velocity components, $u_{\sub\parallel}$ and $u_{\sub\perp}$, are defined as
\beq
\begin{pmatrix}
u_{\sub\parallel} \\
u_{\sub\perp}
\end{pmatrix}
=
\begin{pmatrix}
\cos (\theta) &  \sin (\theta)  \\
\sin (\theta) & -\cos (\theta) 
\end{pmatrix}
\begin{pmatrix}
u^o \\
v^o
\end{pmatrix}
,\ \quad 
\theta = \arg (\tvec) .
\eeq
This rotation is applied to each random realization of the model outputs $\uvec^o$ and $\tau$, without performing daily averaging of the data. We retain the full dataset to compute ensemble statistics before applying low-pass filtering or time-averaging for visualization.

\begin{figure}[htbp]
\centering
\includegraphics[width=\textwidth]{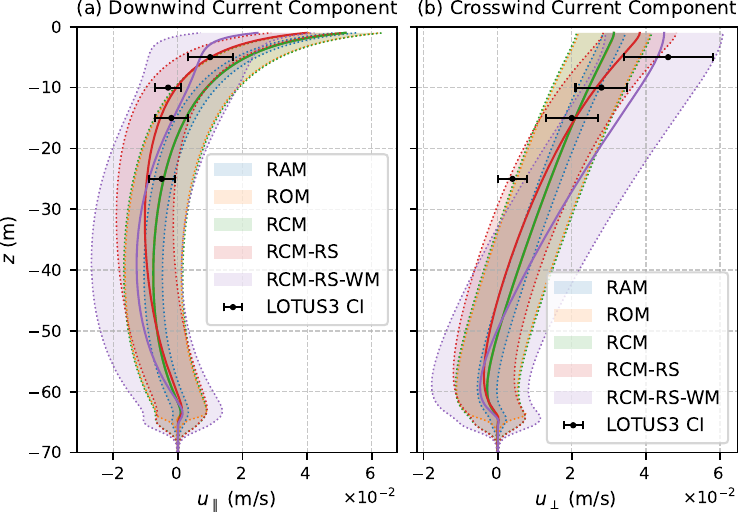}
\caption{Comparison of the ensemble mean (solid lines) and spread (shaded areas) for (a) downwind \LL{(Eulerian)} Ekman current and (b) crosswind \LL{(Eulerian)} Ekman current across different random models. These are compared to the confidence interval (CI, represented by black error bars) of the mean (centered point) LOTUS3 observations (Table 1 of \cite{Price1987wind}) at near-surface depths. The ensemble mean and spread ($\mr{mean} (u) \pm \mr{std} (u)$) are time-averaged over the last 10 days.}
\label{fig:profile}
\end{figure}

%
Fig.~\ref{fig:profile} qualitatively compares the ensemble mean and spread (defined as mean $\pm$ standard deviation) of the Ekman current profiles for different random models, alongside the mean and confidence intervals (CIs) of the observations. It shows that the RAM, which introduces explicit stochastic transport only in the atmosphere, produces a very low ensemble spread and fails to both encompass the mean of the observations and overlap with their CIs over depth. The ROM and RCM, which incorporate explicit stochastic transport in the ocean, increase the spread but still fail to align with the observations near the surface, particularly at $z=(-5, -10)$ m. The RCM-RS and RCM-RS-WM, which include the Coriolis--Stokes force, modify the shape of the current profiles and overlap most of the observation CIs across depth, particularly near the surface. Compared to the other random models, the RCM-RS-WM, which includes additional wave mixing effects, significantly increases the spread over depth and shifts the spread to the left for the downwind component and to the right for the crosswind component.

%
To further evaluate the models’ performance against the LOTUS observations, we employ statistical metrics that account for observational uncertainties. Specifically, we generate observation samples following normal distributions,
$u_{\sub\parallel}^{\sub\mr{obs}} \sim \mc{N} \left( \wh{\mu}_{\sub\parallel}, \wh{\sigma}_{\sub\parallel}^2 \right)$ and $u_{\sub\perp}^{\sub\mr{obs}} \sim \mc{N} \left( \wh{\mu}_{\sub\perp}, \wh{\sigma}_{\sub\perp}^2 \right)$, where the mean values $\wh{\mu}$ and the standard errors $\wh{s} = \alpha \wh{\sigma} / \sqrt{n}$ are obtained from Table 1 of \citet{Price1987wind}. The CI is given by $\mr{CI} = [\wh{\mu}-\wh{s},\wh{\mu}+\wh{s}]$, where $n=53$  represents the effective degrees of freedom used to estimate the standard error over the 160-day record. The confidence levels are set to 95\% ($\alpha = 2$) for the downwind component and 90\% ($\alpha = 1.7$) for the crosswind component. The corresponding sample standard deviations, $\wh{\sigma}_{\sub\parallel}$ and $\wh{\sigma}_{\sub\perp}$, are derived from these values.

We employ two statistical metrics for the model \LL{validation against LOTUS observations}. The first is the Wasserstein distance \citep{Villani2009wasserstein}, which quantifies the similarity between two probability distributions. Specifically, we use the empirical cumulative distribution functions, $\wh{F}_{X_e}$ and $\wh{F}_{X_o}$, of the model ensemble $X_e$ and the randomly generated observations $X_o$, respectively, to compute the $p$-th order Wasserstein distance estimator:
\beq
\wh{W}_p (X_e, X_o) = \left( \int_{\mb{R}} \big| \wh{F}_{X_e} (x) - \wh{F}_{X_o} (x) \big|^p\, \df x \right)^{1/p} .
\eeq
The second metric is the continuous ranked probability score (CRPS) \citep{Weigel2011ensemble}, which evaluates ensemble forecast skill by measuring the integrated squared difference between the cumulative forecast and the observation. In this work, for each realization of the random observations, $x_o^{(i)},\ i = 1,\ldots,n$, the CRPS is computed as
\beqs
\beq
\mr{CRPS} \left( \wh{F}_{X_e}, x_o^{(i)} \right) = \int_{\mb{R}} \left( \wh{F}_{X_e} (x) - \mr{H} \big( x - x_o^{(i)} \big) \right)^2\, \df x ,
\eeq
where $\mr{H}$ is the Heaviside step function. The final CRPS score is obtained by averaging over all sampled observations: 
\beq
\ol{\mr{CRPS}} = \frac{1}{n} \sum_{i=1}^{n} \mr{CRPS} \left( \wh{F}_{X_e}, x_o^{(i)} \right) .
\eeq
\eeqs
To compute these metrics, we independently draw observation samples (following the normal distributions described above) for each time step and depth. The scores are then averaged over depth, yielding time series of the scores for each model. A lower score indicates better model performance.

\begin{figure}[htbp]
\centering
\includegraphics[width=\textwidth]{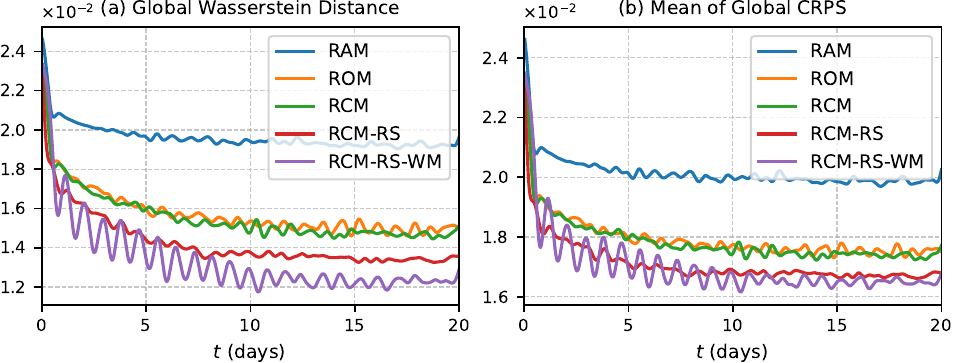}
\caption{Comparison of (a) the globally integrated Wasserstein distance and (b) the mean globally integrated continuous ranked probability score (CRPS) over time for different model ensembles against LOTUS3 observations. Observation samples are drawn from a normal distribution with empirical mean and standard deviation derived from Table 1 of \citet{Price1987wind}. At each time, the CRPS is computed using an observation realization and averaged over samples. A half-day low-pass filter is applied to these time series.}
\label{fig:score}
\end{figure}

Fig.~\ref{fig:score} presents the temporal evolution of the globally integrated first-order ($p=1$) Wasserstein distance and the mean globally integrated CRPS for each model ensemble against LOTUS observations. The results clearly demonstrate an increasing order of model performance:
\[
\text{RAM} < \text{ROM} < \text{RCM} < \text{RCM-RS} < \text{RCM-RS-WM} .
\]
This ranking quantifies the contribution of each additional modeling term in the fully coupled stochastic system to the overall improvement of the model.

\subsection{Statistical diagnostics of energy budget and wind work}\label{sec:stats}
\begin{figure}[htbp]
\centering
\includegraphics[width=\textwidth]{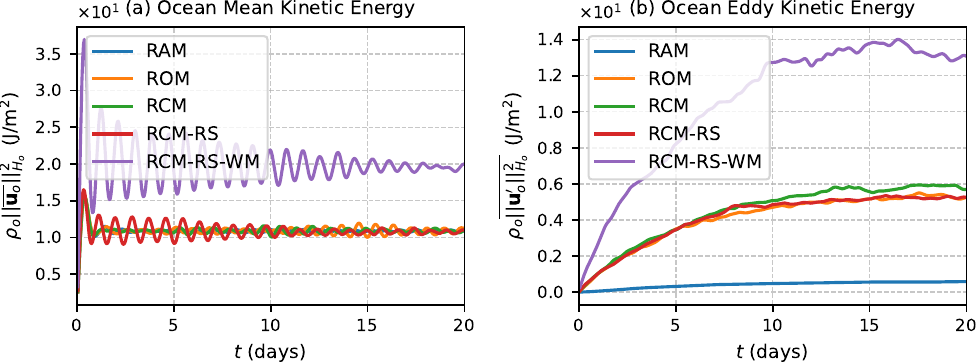}
\par\medskip
\includegraphics[width=\textwidth]{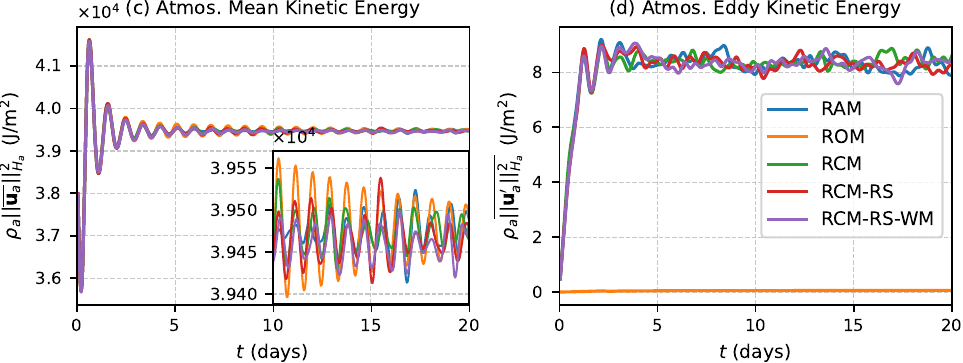}
\caption{Comparison of globally integrated mean kinetic energy (MKE, left) and eddy kinetic energy (EKE, right) for the ocean (top) and atmosphere (bottom) components across different random models (distinguished by color). MKE and EKE are defined in an ensemble sense as $\ol{\uvec} := \wh{\Exp} [\uvec]$ and $\uvec' := \uvec - \wh{\Exp} [\uvec]$. A half-day low-pass filter is applied to these time series. A zoomed-in view of atmospheric MKE over the last 10 days is included to better highlight the differences.}
\label{fig:energy}
\end{figure}

%
%
We begin by analyzing the ensemble decomposition of global energy for both atmospheric and oceanic components, specifically the mean kinetic energy (MKE) and eddy kinetic energy (EKE). The MKE is defined as $\mr{MKE}^\alpha = \rho^\alpha \|\ol{\uvec^\alpha}\|^2$,
while the EKE is given by $\mr{EKE}^\alpha = \rho^\alpha \ol{\|\uvec^\alpha - \ol{\uvec^\alpha}\|^2}$, where $\ol{u}$ denotes the ensemble mean of $u$. Fig.~\ref{fig:energy}(c) shows that all models exhibit similar time-mean values of $\mr{MKE}^a$, albeit with varying levels of oscillation around their respective means. In contrast, Fig.~\ref{fig:energy}(a) demonstrates that the RCM-RS-WM significantly increases the $\mr{MKE}^o$ compared to all the other random models, due to the inclusion of additional wave mixing terms.

As illustrated in Fig.~\ref{fig:energy}(b), RAM, which introduces explicit stochastic transport only in the atmosphere, generates high $\mr{EKE}^a$ but very low $\mr{EKE}^o$. Conversely, as shown in Fig.~\ref{fig:energy}(d), ROM, which introduces explicit stochastic transport only in the ocean, produces high $\mr{EKE}^o$ but very low $\mr{EKE}^a$. The RCM model, which incorporates explicit stochastic transport in both the atmosphere and ocean, yields $\mr{EKE}^o$ levels comparable to ROM and $\mr{EKE}^a$ levels similar to RAM. These results clearly indicate that explicitly representing uncertainties in both components plays a crucial role in determining the variance levels of a coupled system. Furthermore, the two wave-dependent models exhibit $\mr{EKE}^a$ levels similar to those of RCM (Fig.~\ref{fig:energy}d); however, the RCM-RS model slightly reduces $\mr{EKE}^o$, whereas the RCM-RS-WM model significantly enhances it (Fig.~\ref{fig:energy}c).

\begin{figure}[htbp]
\centering
\includegraphics[width=\textwidth]{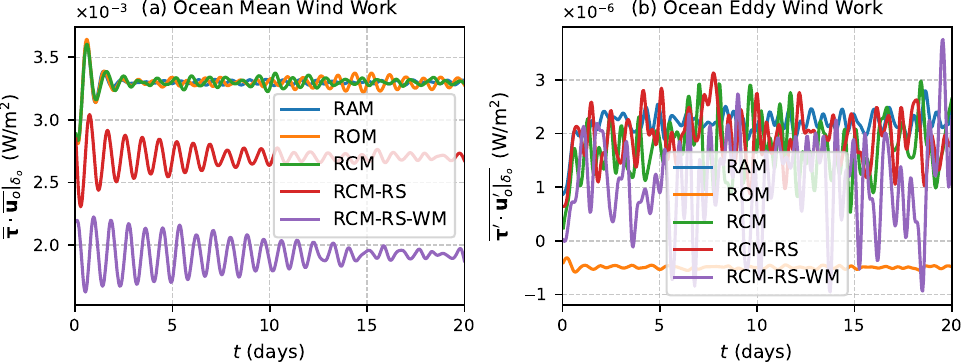}
\par\medskip
\includegraphics[width=\textwidth]{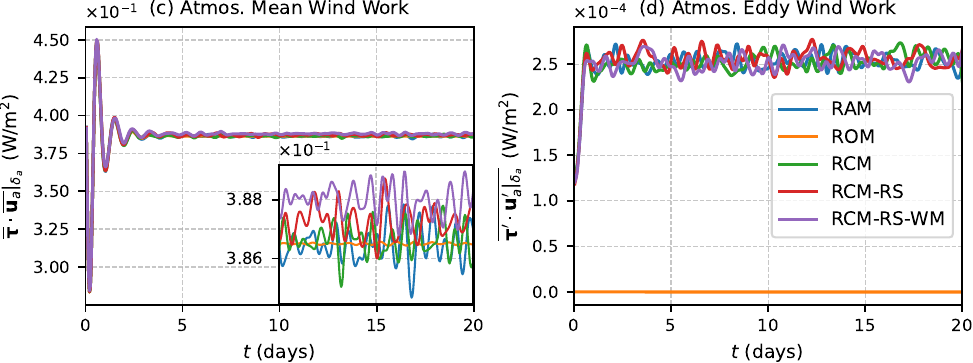}
\caption{Comparison of mean wind work (left) and eddy wind work (right) for the ocean (top) and atmosphere (bottom) components across different random models (distinguished by color).}
\label{fig:work}
\end{figure}

%
We next examine the wind contribution to the ensemble energy budget for both the atmosphere and ocean, characterized by the mean wind work $\mr{MWW}^\alpha = \ol{\tvec} \cdot \ol{\uvec^\alpha}(\delta^\alpha)$ and the eddy wind work $\mr{EWW}^\alpha = \ol{(\tvec - \ol{\tvec}) \cdot (\uvec^\alpha - \ol{\uvec^\alpha}) (\delta^\alpha)}$. Fig.~\ref{fig:work}(a) indicates that surface waves significantly reduce the mean level of $\mr{MWW}^o$ over time while increasing its oscillation, as observed in the comparison of RCM-RS and RCM-RS-WM with RAM, ROM, and RCM. Furthermore, due to wave mixing effects, the reduction in $\mr{MWW}^o$ is more pronounced in RCM-RS-WM than in RCM-RS. Conversely, Fig.~\ref{fig:work}(c) illustrates that surface waves induce a slight increase in the mean level of $\mr{MWW}^a$, an effect that becomes more evident when wave mixing terms are included.

As expected, ROM, which lacks explicit atmospheric uncertainty representation, exhibits an almost negligible $\mr{EWW}^a$ and a small negative $\mr{EWW}^o$, as shown in Figures~\ref{fig:work}(b) and (d). The other models exhibit similar levels of $\mr{EWW}^a$ (Fig.~\ref{fig:work}d). Additionally, the inclusion of wave mixing terms reduces the mean level of $\mr{EWW}^o$ over time while increasing its temporal variability (Fig.~\ref{fig:work}b). 

\subsection{Discussion of wave-dependent surface roughness effects}\label{sec:wave}
We now examine the results obtained using the wave-dependent parameterization of surface roughness length for RCM-RS and RCM-RS-WM. Specifically, we focus on their impact on model performance relative to observations, employing the same statistical metrics as those presented in Section~\ref{sec:lotus}.  

\begin{figure}[htbp]
\centering
\includegraphics[width=\textwidth]{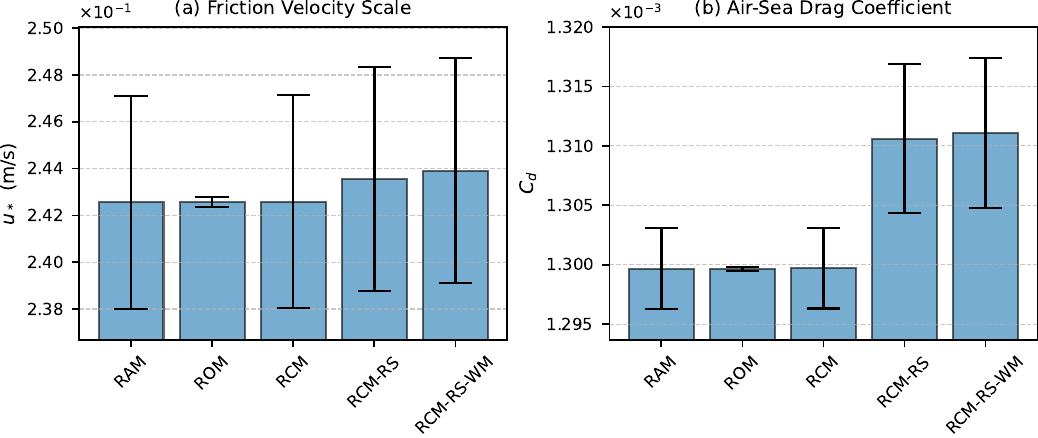}
\caption{Same as Fig.~\ref{fig:diag}, but using the wave age-based roughness parameterization \eqref{eq:wave-age} for RCM-RS and RCM-RS-WM.}
\label{fig:diag-wa}
\end{figure}

\begin{figure}[htbp]
\centering
\includegraphics[width=\textwidth]{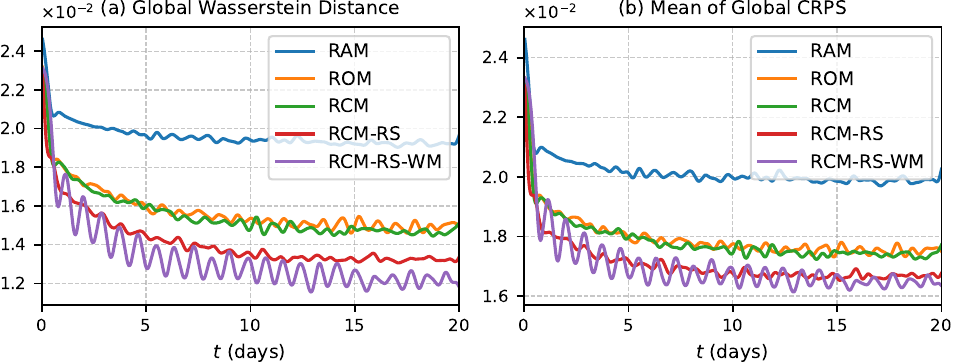}
\caption{Same as Fig.~\ref{fig:score}, but using the wave age-based roughness parameterization \eqref{eq:wave-age} for RCM-RS and RCM-RS-WM.}
\label{fig:score-wa}
\end{figure}

When adopting the wave age-dependent formulation \eqref{eq:wave-age}, Fig.~\ref{fig:diag-wa} illustrates that both the mean friction velocity and air--sea transfer coefficients increase compared to those obtained using the wind speed-dependent formulation \eqref{eq:wind-speed}. In this case, Fig.~\ref{fig:score-wa} demonstrates that the model performance ranking established in Section~\ref{sec:lotus},
\[
\text{RAM} < \text{ROM} < \text{RCM} < \text{RCM-RS} < \text{RCM-RS-WM} ,
\]
remains valid in terms of both the Wasserstein distance and CRPS. However, the relative improvement of RCM-RS-WM over RCM-RS is slightly less pronounced. 
%
%

\begin{figure}[htbp]
\centering
\includegraphics[width=\textwidth]{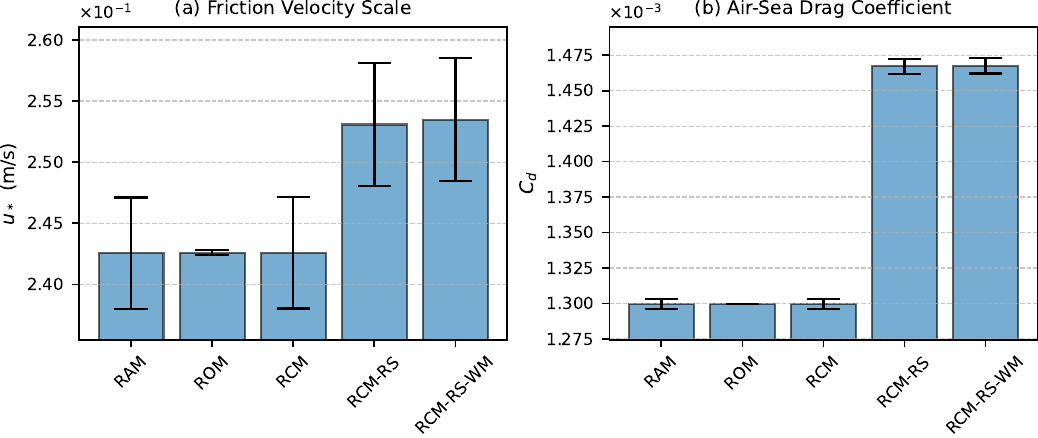}
\caption{Same as Fig.~\ref{fig:diag}, but using the wave age-based roughness parameterization \eqref{eq:sea-state} for RCM-RS and RCM-RS-WM.}
\label{fig:diag-wa-ss}
\end{figure}

For the wave slope-dependent formulation \eqref{eq:sea-state}, Fig.~\ref{fig:diag-wa-ss} indicates a more significant increase in air--sea fluxes compared to the wave age-based case. The model performance ranking remains consistent in this scenario, at least in terms of the Wasserstein distance. However, the improvement of RCM-RS-WM relative to RCM-RS is noticeably less pronounced than in the previous cases, suggesting that stronger air--sea fluxes diminish the relative contribution of wave mixing terms.

\LL{
Throughout this study, all model configurations rely on a common set of turbulence-related parameters---such as those in the KPP and COARE, formulations. This design choice facilitates fair comparisons, but may also bias performance assessments, particularly for models with added physical processes like wave-driven mixing (RCM-RS-WM). In such cases, model–observation mismatch may reflect parameter non-optimality rather than model structure alone. Moreover, the physical interpretation of these parameters may vary across models with distinct dynamics (e.g., inclusion of Stokes shear). A more consistent parameter tuning strategy, such as that proposed in \citet{McWilliams2012wavy}, would help adapt parameterizations to specific model assumptions. Future work may benefit from calibration against high-resolution large eddy simulation (LES) datasets, which provide full-depth current profiles, in contrast to the discrete-depth measurements available in the LOTUS experiment.
}
%
%
\begin{figure}[htbp]
\centering
\includegraphics[width=\textwidth]{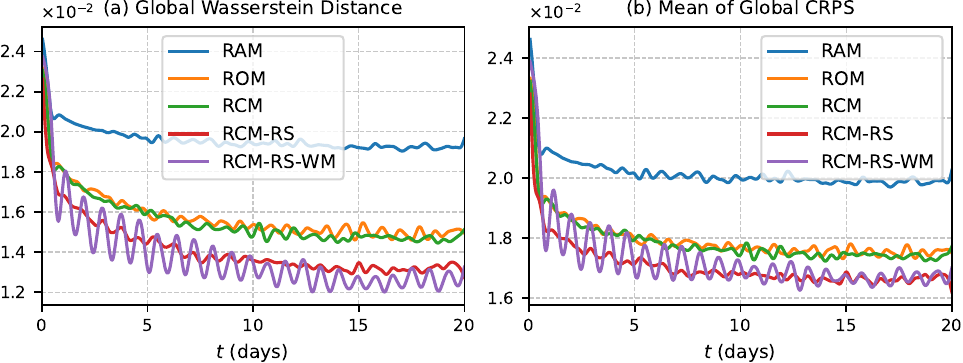}
\caption{Same as Fig.~\ref{fig:score}, but using the wave age-based roughness parameterization \eqref{eq:sea-state} for RCM-RS and RCM-RS-WM.}
\label{fig:score-wa-ss}
\end{figure}


\section{Conclusion}\label{sec:conclu}
We have introduced a stochastic coupled Ekman--Stokes model (SCESM) to represent the atmosphere--ocean boundary layers while incorporating surface wave effects and turbulent fluctuations arising from unresolved scales. The SCESM, along with established parameterizations for air--sea fluxes, turbulent viscosity, and Stokes drift, was rigorously tested through ensemble simulations. Its performance was evaluated against LOTUS data using statistical metrics that account for observational uncertainties.

A performance ranking analysis quantified the impact of various modeling components within the fully coupled stochastic system. The results underscore the importance of explicitly representing uncertainties in both oceanic and atmospheric components to accurately capture the variance levels of the coupled system. Furthermore, while wave-dependent parameterizations of surface roughness enhance air--sea fluxes, they reduce the contribution of wave-induced mixing terms in the SCESM.

\LL{
An important limitation of the present framework lies in the use of classical bulk flux parameterizations, which were originally designed for deterministic, time-averaged conditions. Future developments should consider more consistent formulations of surface coupling that explicitly incorporate uncertainty. Moreover, combining this framework with data assimilation and parameter estimation techniques would further enhance its predictive capability and realism.

Another limitation of the present validation is the lack of direct flux measurements in the LOTUS dataset. While comparisons of mean current profiles serve as a useful benchmark, they offer only indirect constraints on the underlying turbulent exchange mechanisms. More robust evaluation could be achieved using high-resolution coupled LES data \citep[e.g.,][]{Sullivan2025investigation} or modern observational platforms equipped with flux towers and wave-following buoys that measure momentum and heat fluxes.

Beyond these immediate improvements, several extensions are promising: incorporating buoyancy evolution via stochastic equations for stratified Ekman layers with time-varying stratification \citep{McWilliams2009buoyancy}; introducing memory effects through second-order turbulence closures that solve a prognostic TKE equation including surface wave influences \citep{Harcourt2013second}; 
and implementing a spectral representation of surface waves with wave-induced momentum source terms \citep{Perrie2003impact}. 
Together, these developments could significantly advance our ability to simulate and predict coupled \LL{ocean--wave--atmosphere} dynamics.
}

\section*{Acknowledgments}

\LL{
The authors sincerely thank the two reviewers, Baylor Fox-Kemper and Charles Pelletier, for their insightful comments and suggestions, which have greatly helped improve the manuscript. We also acknowledge the support of the ERC EU project 856408-STUOD, which made this research possible.
}


\section*{Data availability statement}

The PyTorch code \textbf{SCESM.py} used to reproduce the simulation data, along with the Python script \textbf{diag.py} which performs all diagnostics from the output data, and the Jupyter notebook \textbf{fig\_stuod.ipynb} which generates the figures presented in this study, are available at \url{https://github.com/matlong/SCESM}.


\bibliographystyle{abbrvnat}
\bibliography{references}

\end{document}